\documentclass[pra,twocolumn,superscriptaddress]{revtex4-1}

\usepackage{braket}
\usepackage{amsmath,bbm}
\usepackage[colorlinks,linkcolor=red,anchorcolor=blue,citecolor=blue]{hyperref}
\usepackage{graphicx}
\usepackage{epstopdf}
\usepackage{dcolumn}
\usepackage{bm}
\usepackage{verbatim}
\usepackage{amsfonts}
\usepackage{color}

\usepackage{hyperref}
\usepackage{float}
\usepackage{amssymb,mathrsfs}

\newcommand{\tr}{{\rm Tr}}
\newcommand{\idol}{\ensuremath{\mathbbm 1}}

\begin{document}

\title{Detecting genuine multipartite entanglement via machine learning}
\author{Yi-Jun Luo} 
\affiliation{School of Physical Science and Technology, Ningbo University, Ningbo 315211, China}
\author{Jin-Ming Liu}
\affiliation{State Key Laboratory of Precision Spectroscopy, School of Physics and Electronic Science,
East China Normal University, Shanghai 200241, China}
\author{Chengjie Zhang}
\email{chengjie.zhang@gmail.com}
\affiliation{School of Physical Science and Technology, Ningbo University, Ningbo 315211, China}
\affiliation{State Key Laboratory of Precision Spectroscopy, School of Physics and Electronic Science,
East China Normal University, Shanghai 200241, China}

\begin{abstract}
In recent years, supervised and semi-supervised machine learning methods such as neural networks, support vector machines (SVM), and semi-supervised support vector machines (S4VM) have been widely used in quantum entanglement and quantum steering verification problems. However, few studies have focused on detecting genuine multipartite entanglement based on machine learning. Here, we investigate supervised and semi-supervised machine learning for detecting genuine multipartite entanglement of three-qubit states. We randomly generate three-qubit density matrices, and train an SVM for the detection of genuine multipartite entangled states. Moreover, we improve the training method of S4VM, which optimizes the grouping of prediction samples and then performs iterative predictions. Through numerical simulation, it is confirmed that this method can significantly improve the prediction accuracy.
\end{abstract}
\date{\today}
\maketitle


\section{Introduction}\label{sec:leve1}
Genuine multipartite entanglement (GME) is a relevant
resource in quantum information processing
\cite{rev1,rev2,rev3,rev4,rev5,rev6}. It is used in many quantum
information tasks, such as cluster states in the one-way quantum computing model \cite{cluster}, Greenberger-Horne-Zeilinger (GHZ) and Dicke states in quantum metrology \cite{Dicke1,Dicke2}, or graph states in quantum error correction codes \cite{QECC1,QECC2}. Consequently, the GME certification is a central task
in the field of quantum information.

There are many entanglement criteria and entanglement measures for bipartite quantum states \cite{rev1,rev2,rev3,rev4,rev5,rev6}, such as the negativity and its extensions \cite{ppt,ppt1,negativity,negativity21,CREN}, the concurrence \cite{EOFde1,2qubit1,2qubit2,concurrence3,concurrence4,concurrence5}, the G-concurrence, \cite{Gour,Fan,Uhlmann,GME1} and the geometric measure of entanglement \cite{GME1,GME2,GME21}, etc.
However, for multipartite quantum systems the situation becomes more complicated, as several different entanglement classes exist \cite{k-entanglement1,k-entanglement2,k-entanglement3,Szalay}. Among all the multipartite entanglement classes, GME can be viewed as the strongest multipartite entanglement type. A multipartite quantum state contains GME if and only if it cannot be expressed as a convex combination of biseparable states with respect to any bipartitions. Many detection criteria and entanglement measures have been proposed for GME \cite{GME3,GME4,GME5,bounds15,Wn,Wn2,bounds16,GME11,GME12,GME13}.

Machine learning is an interdisciplinary field that combines probability theory, statistics, computer science and other domains to study how computers can simulate human learning behavior by constantly reorganizing their existing knowledge structures.
According to the learning style, it is mainly divided into four main categories: supervised learning, unsupervised learning, semi-supervised learning, and reinforcement learning. These methods have been widely used in quantum information, particularly in quantum entanglement classification \cite{lu2018separability,levine2019quantum,mezquita2021review,chen2021detecting,LuH}, quantum steering \cite{zhang2021einstein,Ren}, quantum nonlocality  \cite{broecker2017quantum,canabarro2019machine}, spin system \cite{zhang2017machine,teoh2020machine} and other aspects.

Semi-supervised learning classification algorithms are a form of semi-supervised learning, including the semi-supervised random forest algorithm \cite{leistner2009semi}, safe semi-supervised support vector machine (S4VM) \cite{li2011improving}, semi-supervised k-nearest neighbors algorithm \cite{kramer2013dimensionality} and other algorithms. Among them, S4VM has obvious effects on anomaly detection \cite{montague2017efficient,haweliya2014network} and image text classification \cite{yang2013semi,wan2021semi,sun2022hypergraph}, and has also been applied to quantum steering classification problems \cite{zhang2021einstein}.
By using semi-definite programming (SDP) \cite{cavalcanti2016quantum,wang2020cost,jungnitsch2011taming} to generate quantum entangled states randomly, semi-supervised algorithms can be applied to predict a large number of unlabeled quantum states from a small number of labeled states.

Recently, supervised and semi-supervised machine learning methods such as neural networks, support vector machines (SVM), and S4VM have been widely used in quantum entanglement and quantum steering verification problems.  However, to the best our knowledge, few studies have focused on the detection of GME based on machine learning. Because GME is much more important than bipartite entanglement in many quantum information tasks, and machine learning is a powerful classification tool used for bipartite entanglement, the detection of GME via machine learning is significant and urgent.

In this study, the three-qubit GME detection problem is investigated based on the SVM and S4VM, and these two methods are improved. For the SVM, we use the method of screening favorable support vectors to improve the accuracy of the model while reducing the training time. For the S4VM, based on Ref. \cite{zhang2021einstein}, we propose a new group selection method, which can significantly improve the classification accuracy compared with the direct grouping method. Finally, we compare and analyze the results with SVM, highlighting the superiority of this method. The paper is organized as follows. Section II presents the detection of three-qubit genuine multipartite entanglement using supervised machine learning. In Sec. III, we investigate semi-supervised machine learning for detecting the genuine multipartite entanglement of three-qubit states. Last but not least, Section IV concludes the study.

\section{\label{sec:leve2}SUPERVISED MACHINE LEARNING}
\subsection{\label{sec:leve21}Methods}
An arbitrary three-qubit quantum state $\rho$ can be expressed as
\begin{eqnarray}
	\rho
&=&\frac{1}{8}\bigg({\idol \otimes \idol \otimes \idol+\sum_{i=1}^{3} r_{i} \sigma_{i} \otimes \idol \otimes \idol+\sum_{j=1}^{3} s_{j} \idol \otimes\sigma_{j} \otimes \idol   } \nonumber\\
	&& {\ \ \ \  +  \sum_{k=1}^{3} p_{k} \idol \otimes \idol \otimes \sigma_{k}+\sum_{i, j=1}^{3} t_{i j}\sigma_{i}\otimes \sigma_{j}\otimes\idol  } \nonumber\\
	&& {\ \ \ \  +\sum_{i, k=1}^{3} q_{i k} \sigma_{i} \otimes \idol \otimes \sigma_{k}+\sum_{j, k=1}^{3} o_{j k} \idol \otimes \sigma_{j} \otimes \sigma_{k}} \nonumber\\
	&& {\ \ \ \  +\sum_{i, j, k=1}^{3} m_{i j k} \sigma_{i} \otimes \sigma_{j} \otimes \sigma_{k}}\bigg),
\end{eqnarray}
where $\idol$ is $2 \times 2$ identity matrix and $\sigma_{i}$ $(i=1,2,3)$ are Pauli matrices.

Based on the entanglement witness $W=P_{\alpha}+Q_{\alpha}^{T_{\alpha}}$ for bipartition $\alpha|\bar{\alpha}$ with positive operators $P_{\alpha}$ and $Q_{\alpha}$,
Ref.~\cite{jungnitsch2011taming} defined an entanglement monotone to quantify GME, i.e., the genuine multipartite negativity (GMN), which can be easily computed via SDP. Moreover, Ref.~\cite{Hofmann2014} proposed a renormalized version of the GMN. For a three-qubit state $\rho_{A B C}$, the renormalized GMN $N_g(\rho_{ABC})$ is given by
\begin{equation}
	\begin{aligned}
		& \ \ \ N_g(\rho_{ABC})=  -  \inf \tr(W \rho_{A B C}) \\
		& \text { subject to: }   W=P_{\alpha}+Q_{\alpha}^{T_{\alpha}} \\
		& \ \ \ \ \ \  \ \ \ \ \ \ \ \ \ \ \ \  0\leq P_{\alpha} \\
		& \ \ \ \ \ \  \ \ \ \ \ \ \ \ \ \ \ \  0\leq Q_{\alpha}\leq \idol \text{ for all bipartitions } \alpha|\bar{\alpha},
	\end{aligned}\label{eq.2}
\end{equation}
where $\alpha$ runs over all possible subsystems in $\{A,B,C\}$, and $T_{\alpha}$ is the partial transpose for subsystem $\alpha$. Notably, the renormalized GMN is equal to a mixed-state convex roof of bipartite negativity,
\begin{eqnarray}\label{GMN}
N_g(\varrho)=\inf_{p_\alpha,\varrho_{\alpha}}\sum_\alpha p_\alpha N_{\alpha}(\varrho_{\alpha}),
\end{eqnarray}
where the summation runs over all possible decompositions $\alpha|\overline{\alpha}$ of the system and the minimization is performed over all mixed state decompositions of the state $\varrho=\sum_\alpha p_\alpha \varrho_{\alpha}$.

Using SDP (\ref{eq.2}), we can randomly generate three-qubit quantum states and obtain their renormalized GMN. If the renormalized GMN value is positive, we consider the quantum state to be genuinely entangled and label it as -1. If the target value is non positive, we label it as +1. To balance the data, we created 55,000 quantum states with +1 labels and 55,000 quantum states with -1 labels, randomly classified all quantum states, and finally took 44,000 quantum states with +1 labels and 44,000 quantum states with -1 labels as the training set, and  11,000 quantum states with +1 labels and 11,000 quantum states with -1 labels as the test set to generate a three-qubit quantum entanglement classifier.

The built-in algorithm we used for this quantum classifier is the SVM algorithm, which is an efficient supervised learning method that classifies data by looking for a classification line between two types of data differences. The SVM algorithm was first developed by Cortes and Vapnik in 1995 \cite{cortes1995support}. The basic principle is that if the training data are points distributed on a two-dimensional plane, they are clustered into different regions according to their classification. The classification algorithm based on classification boundaries determines the boundary between these classifications by training various data points, and then obtains the corresponding fitting curve. For many $N$-dimensional data, it can be thought of as points in $N$-dimensional space, and the classification boundary is a polygon in $N$-dimensional space, called a hypersurface (hypersurfaces are one dimension smaller than $N$-dimensional space).

Linear classifiers use the boundaries of hyperplane types, whereas nonlinear classifiers use hypersurfaces. Suppose we have a part of the original data to classify: ($x_{11}$, $x_{12}$, $\cdots$, $x_{1n}$, $y_{1}$), $\cdots$, ($x_{m1}$, $x_{m2}$, $\cdots$, $x_{mn}$, $y_{m}$), where $x_{1}$, $\cdots$, $x_{n}$ represents the feature size of the data in different dimensions, and $y_{m}$ represents the category of this set of data, assuming it is a 2 classification problem, that is +1 or -1. Target hyperplane:
\begin{equation}
	\omega^{T} x+b=0.
\end{equation}
For the optimal $\omega$ and $b$,  the distance between any point in the space ($x_{1}$, $x_{2}$,..., $x_{n}$) to the target hyperplane is
\begin{equation}
r=\frac{\left|\omega^T x+b\right|}{\|\omega\|}.
\end{equation}
We use the marking method as follows,
\begin{equation}
\begin{array}{c}
	\left\{\begin{array}{ll}
		\omega^{T} x_{i}+b \geq 1, & y_{i} = +1 \\
		\omega^{T} x_{i}+b \leq -1, & y_{i} = -1
	\end{array}\right.
\end{array}
\end{equation}
with the target being
\begin{equation}
\begin{gathered}
	\min _\omega \frac{1}{2} \omega^T \omega \\
	\text { such that } y_i\left(\omega x_{i}+b\right) \geq 1.
\end{gathered}
\end{equation}

For nonlinear classifiers, the kernel function method is used. The basic idea is to map the data into a high-dimensional space through a nonlinear transformation (kernel function), make the data linear in it, and finally apply a simple linear SVM for classification. There are many kinds of Kernel functions, such as linear kernels, polynomial kernels, Gaussian kernels, Laplace kernels, and Sigmoid kernels. Here, we used the following Gaussian kernels:
\begin{equation}
\kappa\left(x_{i},x_{j}\right)=\varphi\left(x_{i}\right) \cdot \varphi\left(x_{j}\right)=e^{-\gamma\left\|x_{i}-x_{j}\right\|^2}.
\end{equation}
After introducing relaxation variables $\xi_i$ and kernel functions $\kappa$,
\begin{equation}
\begin{gathered}
	\min _{\omega, b, \xi} \frac{1}{2} \omega^T \omega+C \sum_{i=1}^n \xi_i \\
	\text { such that } y_i\left(\omega \varphi\left(x_{j}\right)+b\right) \geq 1-\xi_i,\\
	\xi_i \geq 0.
\end{gathered}
\end{equation}

\begin{figure}
	\centering
	\includegraphics[width=3.7in]{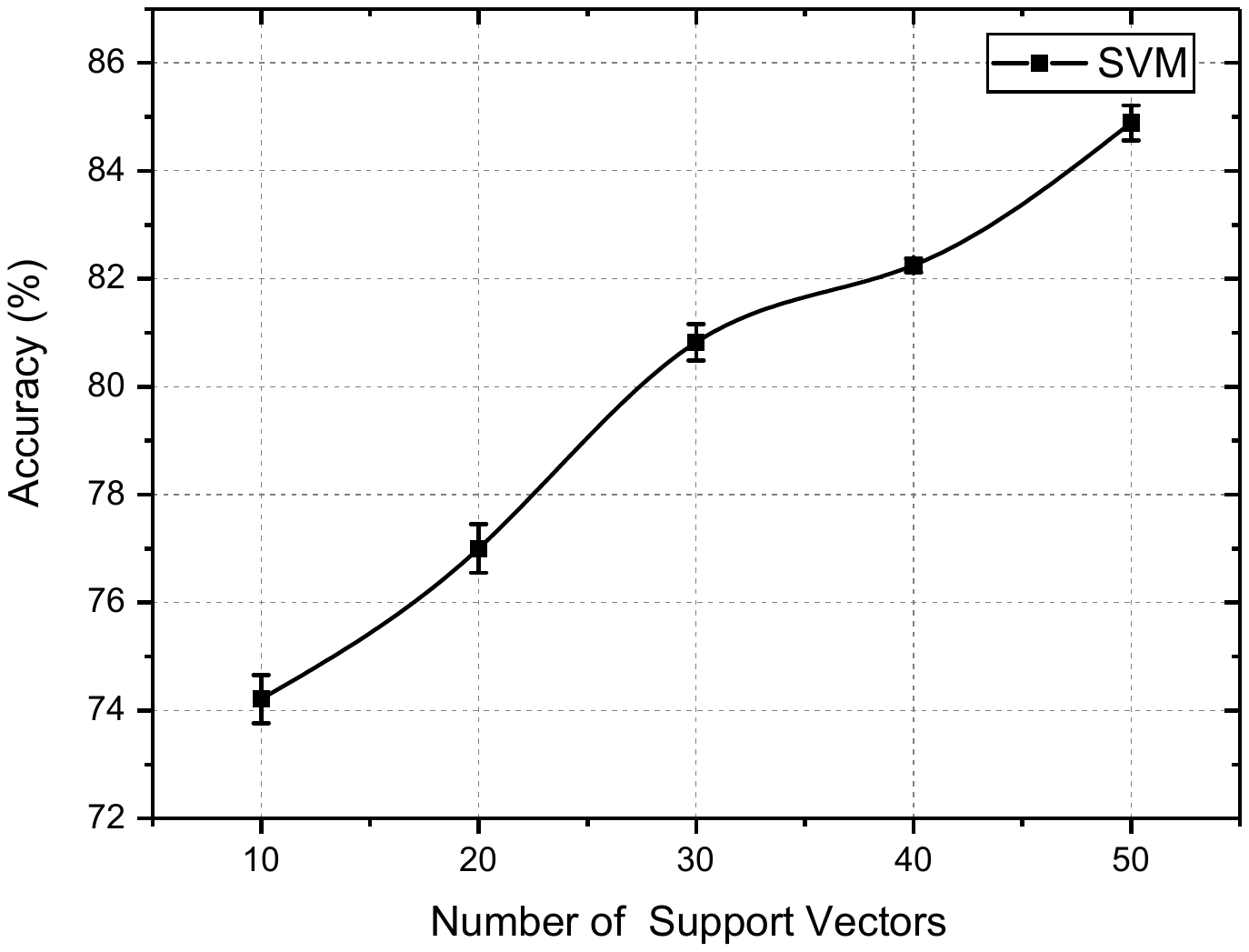}\\
	\caption{The accuracy and the error bar of different numbers of support vectors.}\label{FIG.1}
\end{figure}

The value $C$ is the penalty coefficient or penalty factor, which represents the tolerance of the model to error, and can be understood as the weight of adjusting the preference of the two indicators (interval size and classification accuracy) in the optimization direction, that is, the tolerance of error. The higher $C$, the more intolerant the error is, so it is easy to overfit the model; the smaller $C$, the easier it is to underfit, so whether $C$ is too large or too small, the generalization ability will be worse. $\gamma$ parameters implicitly determine the distribution of the data after mapping to a new feature space. Similarly, being too large or too small reduces classification accuracy. Therefore, when using the SVM algorithm, the selection of parameters $C$ and $\gamma$ is very important, and the choice  directly affects the quality of the final result classification. In this study,  five-fold cross-verification and grid search are adopted to search for optimal values of $C$, $\gamma$ (we obtained $C=3.5, \gamma=-1.8$ by searching), and the search can effectively reduce the training time and avoid finding the local optimal solution.

\subsection{\label{sec:leve22}Improved SVM and Numerical Results}
In the numerical simulation, to meet the needs of our experiment, we first generated a large number of manipulable three-qubit quantum state samples, and obtained the positive and non-positive renormalized GMN values of these quantum state samples through the SDP algorithm, labeled +1 and -1, respectively. The stretching subsystem stretches the random density matrix puts it into the SVM as a feature vector for training, and obtains the prediction accuracy of the corresponding quantum state. Taking a three-qubit quantum state as an example, its density matrix is 8 $\times$ 8, stretched to 64, corresponding to the support vector of SVM. Support vectors that are not conducive to classification are found through traversal search. Thus, the accuracy of SVM classification can be effectively improved. In Fig.~\ref{FIG.1}, different numbers of support vectors are used to obtain the accuracy of unlabeled samples.

\begin{table}
	\centering
	\renewcommand\arraystretch{1.5}
	\tabcolsep=0.425cm
	\caption{Classification accuracy ($\%$) of support vector machine before and after filtering dimensions.}\label{tab.1}
	
	\begin{tabular}{l*{3}c}
		\hline
\hline
		number     &  SVM        &  After dimension reduction  \\\hline
		$110,000$      & 86.80      & 87.67 \\
		\hline
\hline
	\end{tabular}
\end{table}


Through the traversal search, it is found that the 23rd, 36th, 51st, 54th, 55th, 57th, and 63rd dimensions in support vectors of three-qubit quantum states can be reduced for the accuracy. In Table~\ref{tab.1}, it is found that the accuracy after filtering the above dimensions is approximately $1\%$ higher than before. From these results, it can be observed that not all support vectors of quantum states are conducive to the classification of genuinely entangled states. This opens up the possibility of choosing suitable support vectors and finding the rules between them in future research.

\section{\label{sec:leve3}SEMI-SUPERVISED MACHINE LEARNING}
\subsection{\label{sec:leve31}Methods}
In real life, the vast majority of data do not have category labels, and data with category labels is only a small part. In this case, supervised machine learning often does not predict unlabeled data well, because there is too little labeled data in training. Thus, the SVM cannot accurately delineate classification boundaries and can easily affect the accuracy of training. At the same time, supervised methods rely on strong human intervention, and the cost of manual labeling is very high, resulting in the scarcity of manually labeled samples, and a large amount of unlabeled data does not participate in training, making supervised machine learning not highly practical in practical applications. Subsequently, semi-supervised machine learning was developed based on supervised machine learning, which can better solve this problem. Semi-supervised machine learning uses a small number of labeled samples to train a large number of unlabeled samples, which can maximize the use of all data, ensure classification accuracy, and reduce the cost of training.

A semi-supervised support vector machine (S3VM) \cite{cortes1995support} is an early semi-supervised learning method that attempts to normalize and adjust decision boundaries by exploring unlabeled data based on clustering assumptions. However, the objective function of the S3VM algorithm is not convex, it has multiple local optimal solutions. To solve the specific problem of a certain aspect of S3VM, many derivative algorithms have been produced. The meanS3VM \cite{li2009semi} algorithm is proposed to improve the efficiency of S3VM, the S4VM \cite{li2014towards} algorithm is proposed to focus on multiple possible low-density demarcations at the same time, and the CS4VM \cite{li2010cost} algorithm is proposed to improve the cost-sensitive problem.

In our study, the S4VM algorithm is used to classify the entanglement of quantum states and some improvements over the traditional S3VM are achieved. The traditional S3VM is based on the low-density hypothesis, which attempts to find a low-density dividing line, i.e., a low-density region that prefers to make decisions that the boundary passes through the feature space. As we know, in semi-supervised machine learning, there are fewer label data and multiple classification boundaries,  it is impossible to determine which works best. The S3VM focuses on an optimal low-density dividing line, and sometimes the results are not good, while S4VM, focuses on multiple possible low-density dividing lines at the same time, considering more comprehensively than S3VM.

S4VM algorithm:
\begin{equation}
\begin{gathered}
	\min _{\left\{\omega_t, b_t, \hat{y}_t \in B\right\}_{t=1}^T} \sum_{t=1}^T\left(\frac{1}{2} \omega_t^T \omega_t+C_1 \sum_{i=1}^l \xi_i+C_2 \sum_{j=1}^{l} \hat{\xi}_j\right) \\
	+G \sum_{1 \leq t \neq \tilde{t} \leq T} \delta\left(\frac{\hat{y}_t^{\prime} \hat{y}_{\tilde{t}}}{u} \geq 1-\varsigma\right), \\
	\text { such that } y_{i}\left(\omega_{\mathrm{t}}^{\prime} \varphi\left(\mathbf{x}_{i}\right)+\mathrm{b}_{\mathrm{t}}\right) \geq 1-\xi_{i}, \quad \xi_{i} \geq 0 \\
	\qquad\qquad\hat{y}_{t, j}\left(\omega_t^{\prime} \varphi\left(\hat{\mathbf{x}}_j\right)+b_t\right) \geq 1-\hat{\xi}_j, \quad \hat{\xi}_j \geq 0 \\
	\forall i=1, \cdots, l, \quad \forall j=1, \cdots, u, \quad \forall t=1, \cdots, T.
\end{gathered}\nonumber
\end{equation}

Through the recent work \cite{zhang2021einstein}, it can be found that the method of group prediction is adopted for the S4VM algorithm to improve the prediction accuracy. If there are 4000 unlabeled quantum states, they are divided into $M$ groups, and predicted  sequentially.  One can treat the results of each prediction as true labels, and retrain the predictions. This method has a higher accuracy than the S4VM algorithm, and the average accuracies of S4VM for $m=4$ and $m=8$ on the bipartite steering is 0.969 and 0.979 respectively.

\begin{figure}
	\centering
	\includegraphics[width=3.5in]{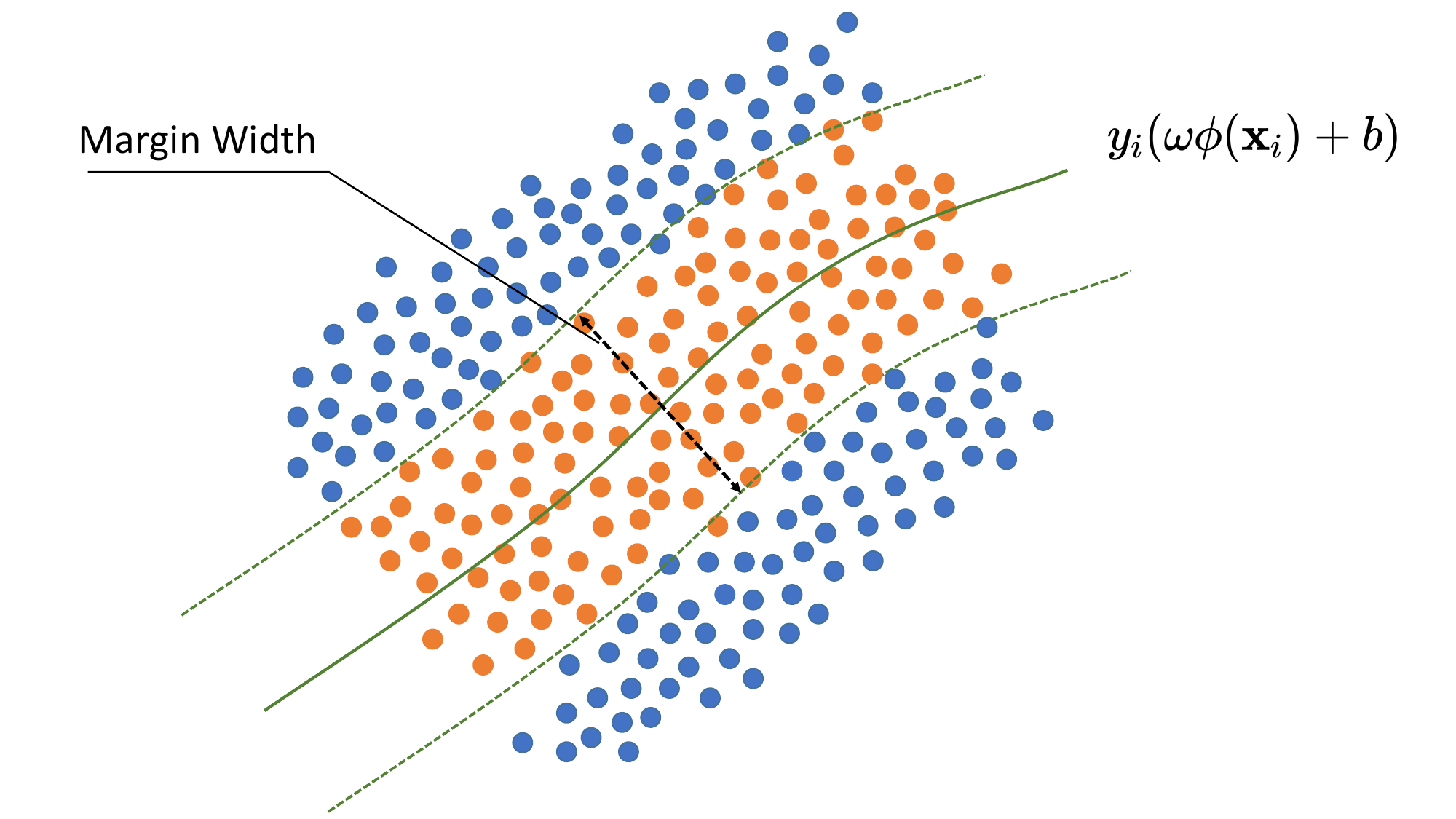}\\
	\caption{Selected S4VM method with $m=2$, where $m$ denotes that we divide all data points into $m$ groups. Blue points are the first group of prediction data points, orange points are the second group of prediction data points.}\label{FIG.2}
\end{figure}

We improved the SVM-S4VM algorithm in Ref. \cite{zhang2021einstein}, named Renewal SVM-S4VM, to make predictions for unlabeled quantum states after selectively grouping. As shown in Fig.~\ref{FIG.2}, the accuracy of the predictions can be significantly improved when the data set is far from the other side of the fitted curve. Therefore, we first use S4VM to sort the distribution of all unlabeled data points, and then group them to make predictions from outside to inside.

The S4VM program used is the MATLAB program of the Institute of Machine Learning and Data Mining of Nanjing University, and the effect of the article has been achieved after corresponding rewriting. $C1$, $C2$, $\gamma$ in the S4VM model are obtained through a grid search and five-fold cross-validation.

Before running the selective S4VM algorithm, $l/2$ three-qubit quantum states with +1 labels and $l/2$ three-qubit quantum states with -1 labels should be prepared by SDP, which are considered as the labeled data. In addition, 1000 quantum states with +1 labels and 1000 quantum states with -1 labels are divided into 2 groups ($m=2$), which are treated as unlabeled data for testing (actually these 2000 states are labeled, we treat them as unlabeled data for testing algorithm).

(1) The S4VM algorithm was applied to 2000 unlabeled quantum states with $l$ labeled quantum states, and the fitted value size of each quantum state was obtained after obtaining the best cross-verification accuracy and hyperparameters, reflecting the degree of entanglement. The data points are arranged from the largest to the smallest, and finally  500 positive maximum data and 500 negative maximum data are taken into the first group, and the rest into the second group.

(2) Taking the $l$ labeled data as the training set, predict the first set of 1000 data, obtaining the best cross-validation accuracy and the best hyperparameters, and treat the prediction results as correct labels.

(3) $l$ raw data and the first set of 1000 labeled data were used as the training set, and the second group of 1000 data was predicted to obtain the best cross-validation accuracy and  hyperparameters.

(4) The average accuracy of the two groups of unlabeled quantum states is regarded as the classification accuracy of the unlabeled quantum states.

\subsection{\label{sec:leve32}Numerical Results}
In the numerical simulation, we generated quantum state samples and manipulate quantum states using MATLAB, and determined whether they were entangled states through SDP algorithms and mark them. To balance the number of samples, we randomly generated the same numbers of ``+1" and ``-1" labeled three-qubit quantum states, where 20 ``+1" labels and 20 ``-1" labels were used as known the samples of S4VM, SVM-S4VM, and Renewal SVM-S4VM semi-supervised learning,  1000 ``+1" labels and 1000 ``-1" labels were used as prediction samples, and  prediction accuracies of $n=2$, 4, 8 and 16 were obtained respectively. These steps were repeated to obtain six sets of training systems consisting of different known labels.

\begin{figure}[t]
	\centering
	\includegraphics[width=3.5in]{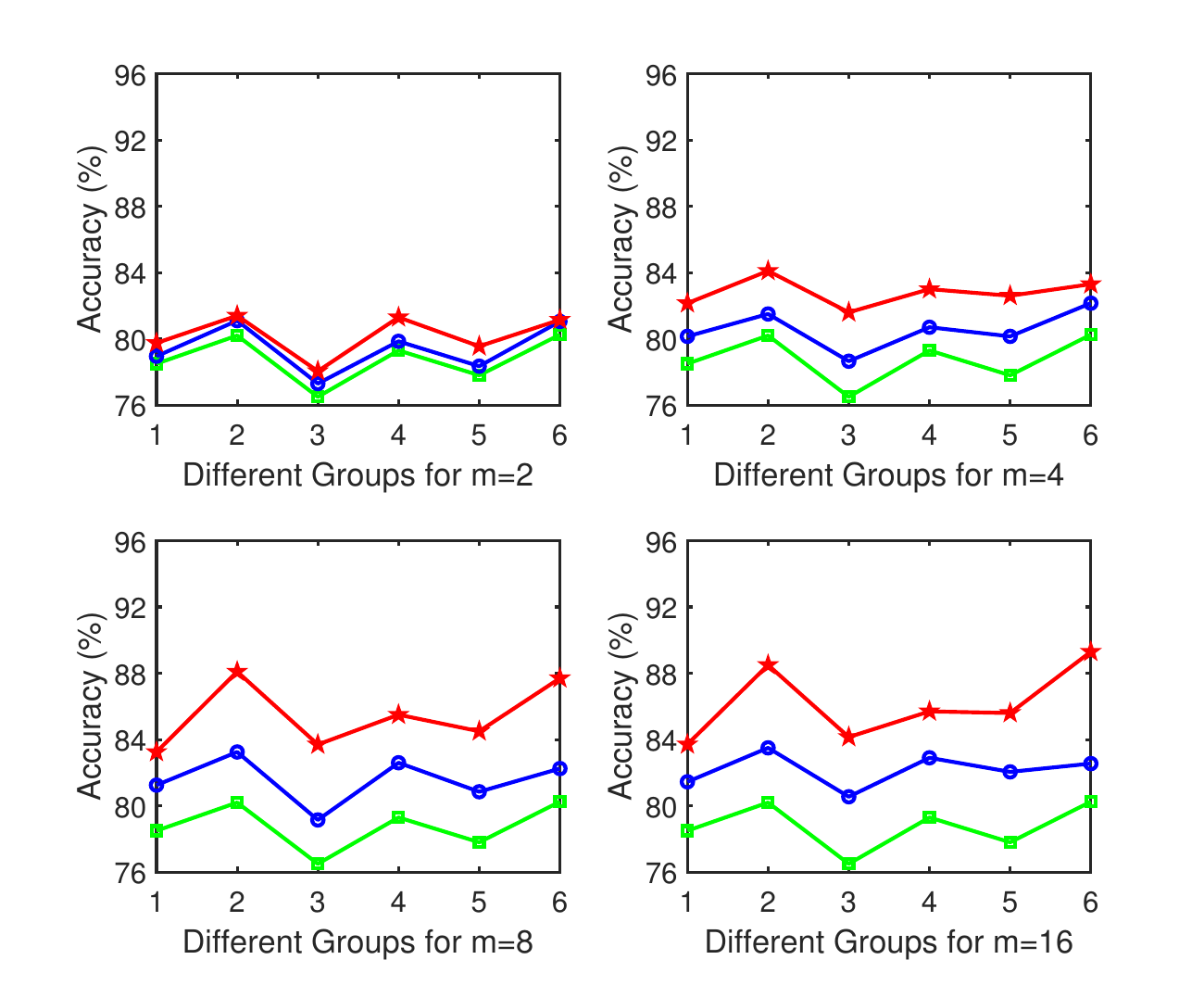}\\
	\caption{When $m=2$, 4, 8, 16, average accuracy ($\%$) of three algorithms under different $m$ for 40 labeled quantum states. The accuracies by the Renewal SVM-S4VM are represented by red lines with pentagram, the accuracies by the SVM-S4VM are represented by blue lines with circle, the accuracies by the S4VM are represented by green lines with square, respectively.}\label{FIG.3}
\end{figure}

When $m=1$ in this paper, the S4VM, SVM-S4VM, and the Renewal SVM-S4VM methods are not substantially different because they do not use the above grouping prediction methods. Based on this, this paper compares the advantages and disadvantages of the three methods for the different cases of $m=2$, 4, 8 and 16.

\begin{figure}[t]
	\centering
	\includegraphics[width=3.5in]{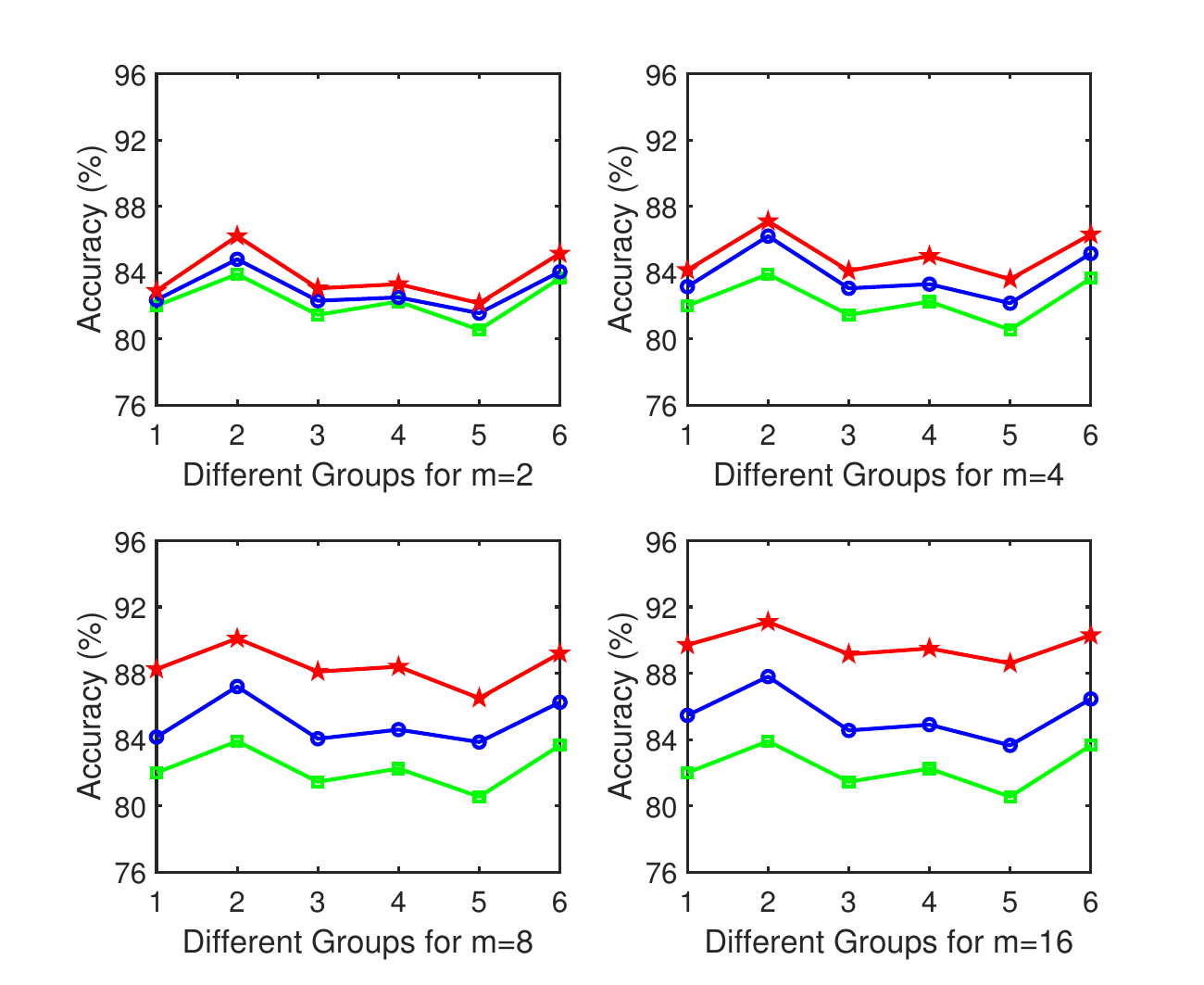}\\
	\caption{When $m=2$, 4, 8, 16, average accuracy ($\%$)  of three algorithms under different $m$ for 60 labeled quantum states. The accuracies by the Renewal SVM-S4VM are represented by red lines with pentagram, the accuracies by the SVM-S4VM are represented by blue lines with circle, the accuracies by the S4VM are represented by green lines with square, respectively.}\label{FIG.4}
\end{figure}	

In Fig.~\ref{FIG.3}, we use six different groups of 40 labeled quantum states to implement SVM-S4VM and Renewal SVM-S4VM, obtain the classification accuracy of 2000 unlabeled quantum states in the case of $m=2$, 4, 8, 16, and compare it with the direct prediction results of S4VM without group prediction. It can be observed from Fig.~\ref{FIG.3} that in the four cases, SVM-S4VM and Renewal SVM-S4VM are larger than S4VM, and the prediction accuracy of Renewal SVM-S4VM is greater than that of SVM-S4VM on the whole. It is obvious that the improvement in the Renewal SVM-S4VM prediction accuracy on SVM-S4VM is more obvious when $m=8$ and 16, followed by $m=4$, and the improvement is the smallest at $m=2$. Except for the sixth group $m=4$, 8, the fourth group $m=8$, 16, etc., SVM-S4VM will hardly improve as $m$ increases, and the prediction accuracy of most SVM-S4VM and Renewal SVM-S4VM will increase with the increasement of $m$.

\begin{figure}[t]
	\centering
	\includegraphics[width=3.5in]{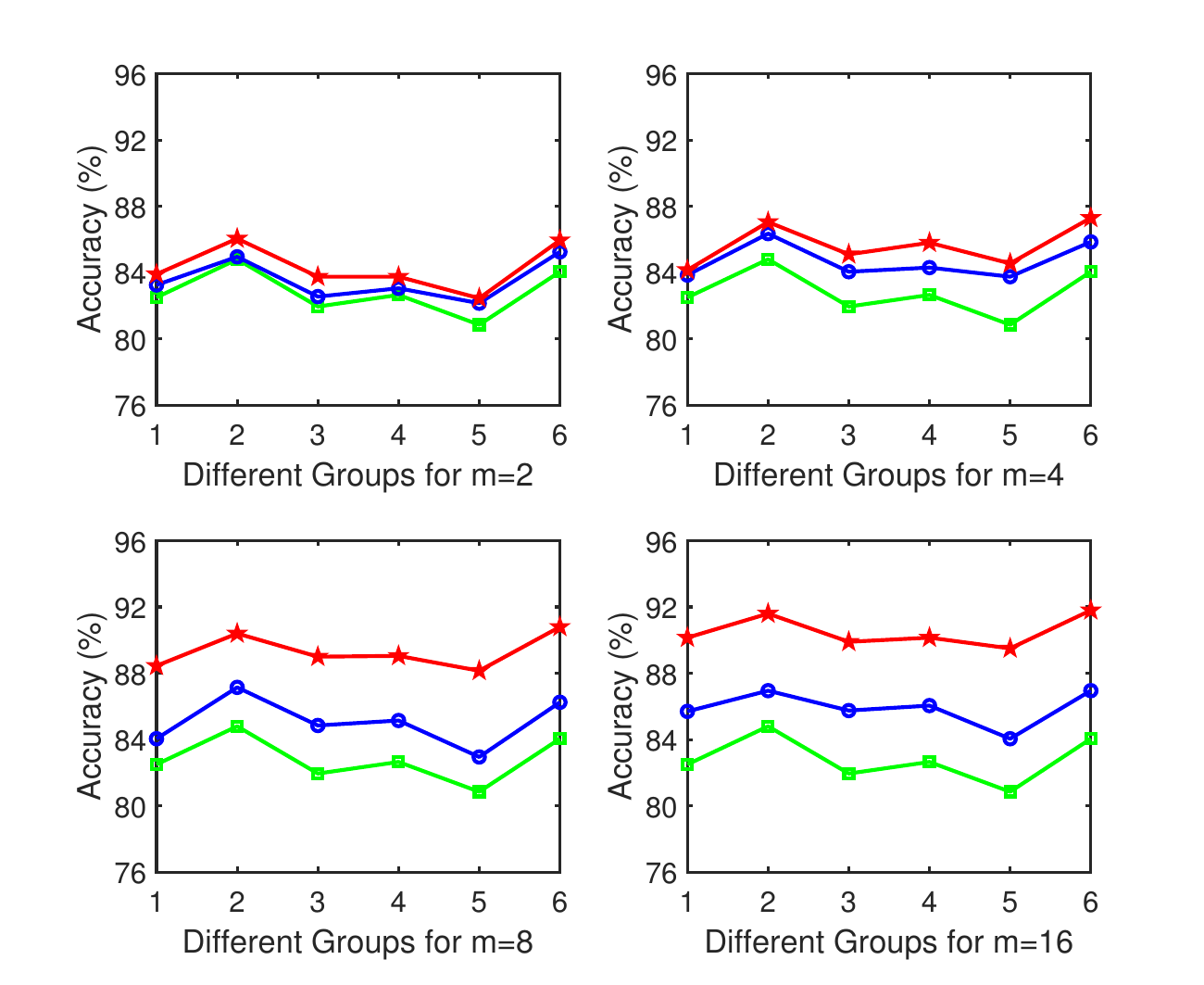}\\
	\caption{When $m=2$, 4, 8, 16, average accuracy ($\%$) of three algorithms under different $m$ for 80 labeled quantum states. The accuracies by the Renewal SVM-S4VM are represented by red lines with pentagram, the accuracies by the SVM-S4VM are represented by blue lines with circle, the accuracies by the S4VM are represented by green lines with square, respectively.}\label{FIG.5}
\end{figure}

\begin{figure}[t]
	\centering
	\includegraphics[width=3.5in]{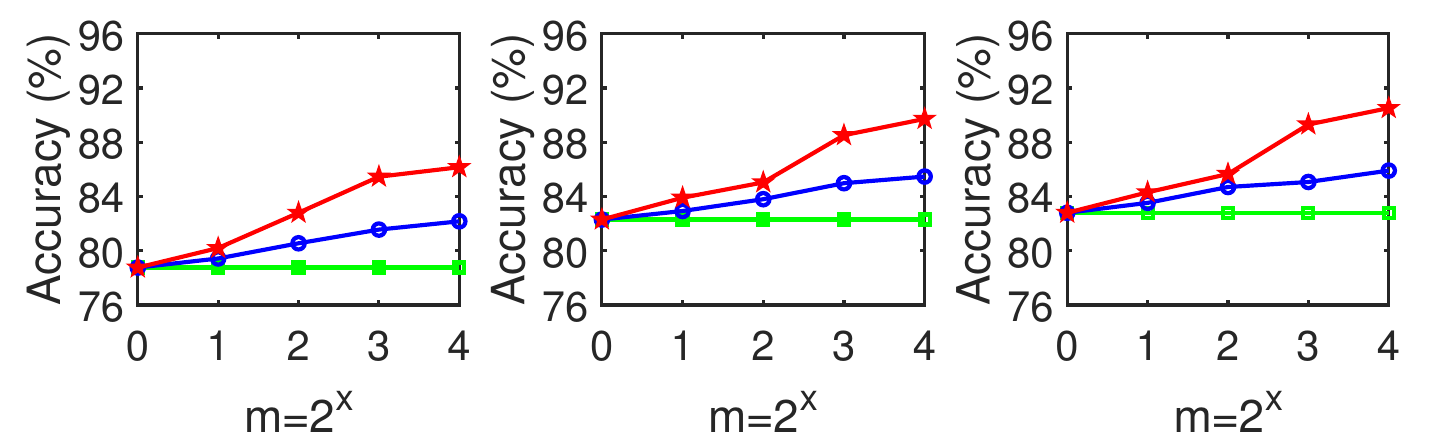}\\
	\caption{Average accuracy ($\%$) of three algorithms under different $l$ for 40, 60, 80 labeled quantum states. The accuracies by the Renewal SVM-S4VM are represented by red lines with pentagram, the accuracies by the SVM-S4VM are represented by blue lines with circle, the accuracies by the S4VM are represented by green lines with square, respectively.}\label{FIG.6}
\end{figure}

\begin{figure*}[t]
	\begin{minipage}[t]{0.32\textwidth}
		\centering
		\includegraphics[width=\textwidth]{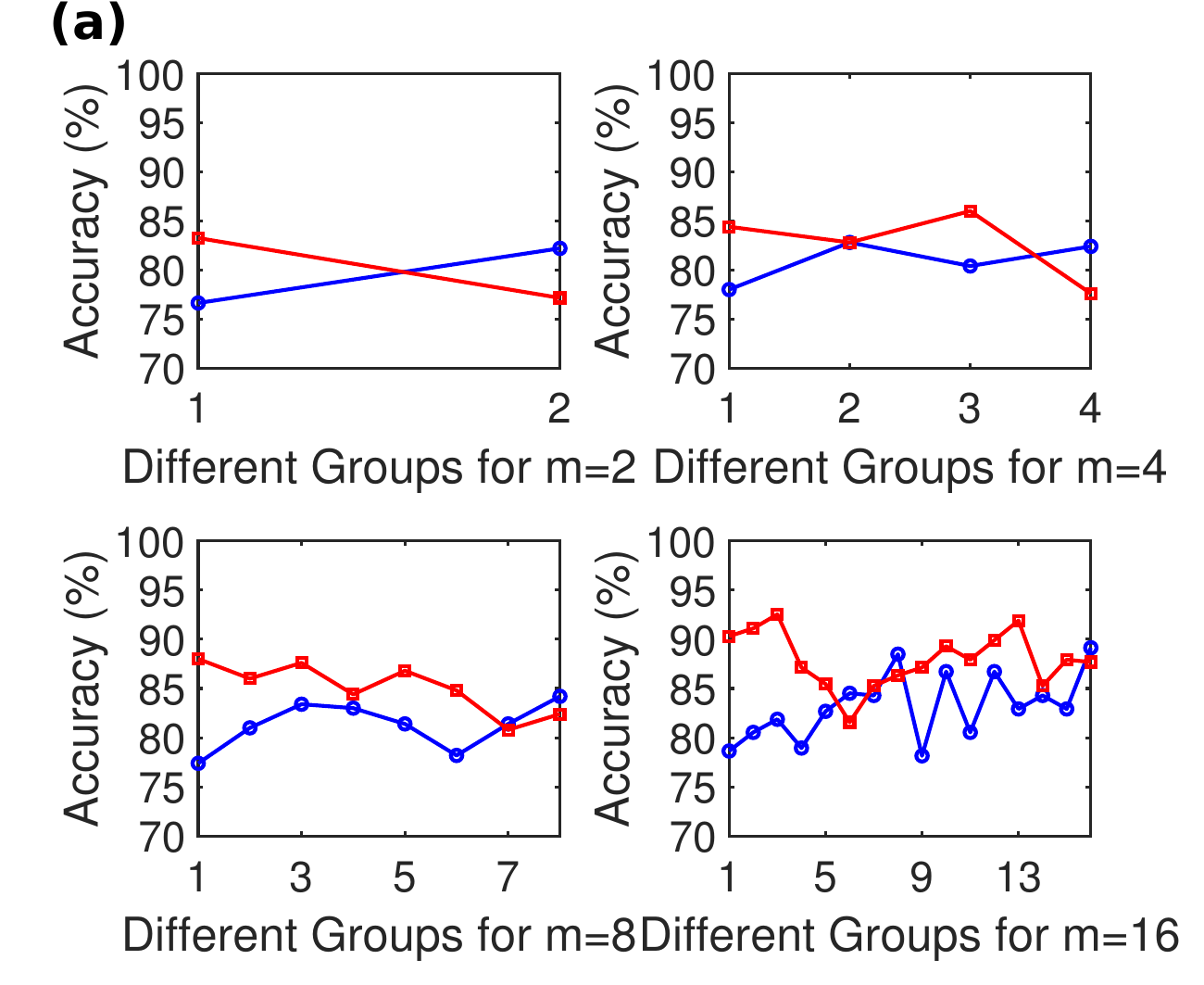}
	\end{minipage}
	\begin{minipage}[t]{0.32\textwidth}
		\centering
		\includegraphics[width=\textwidth]{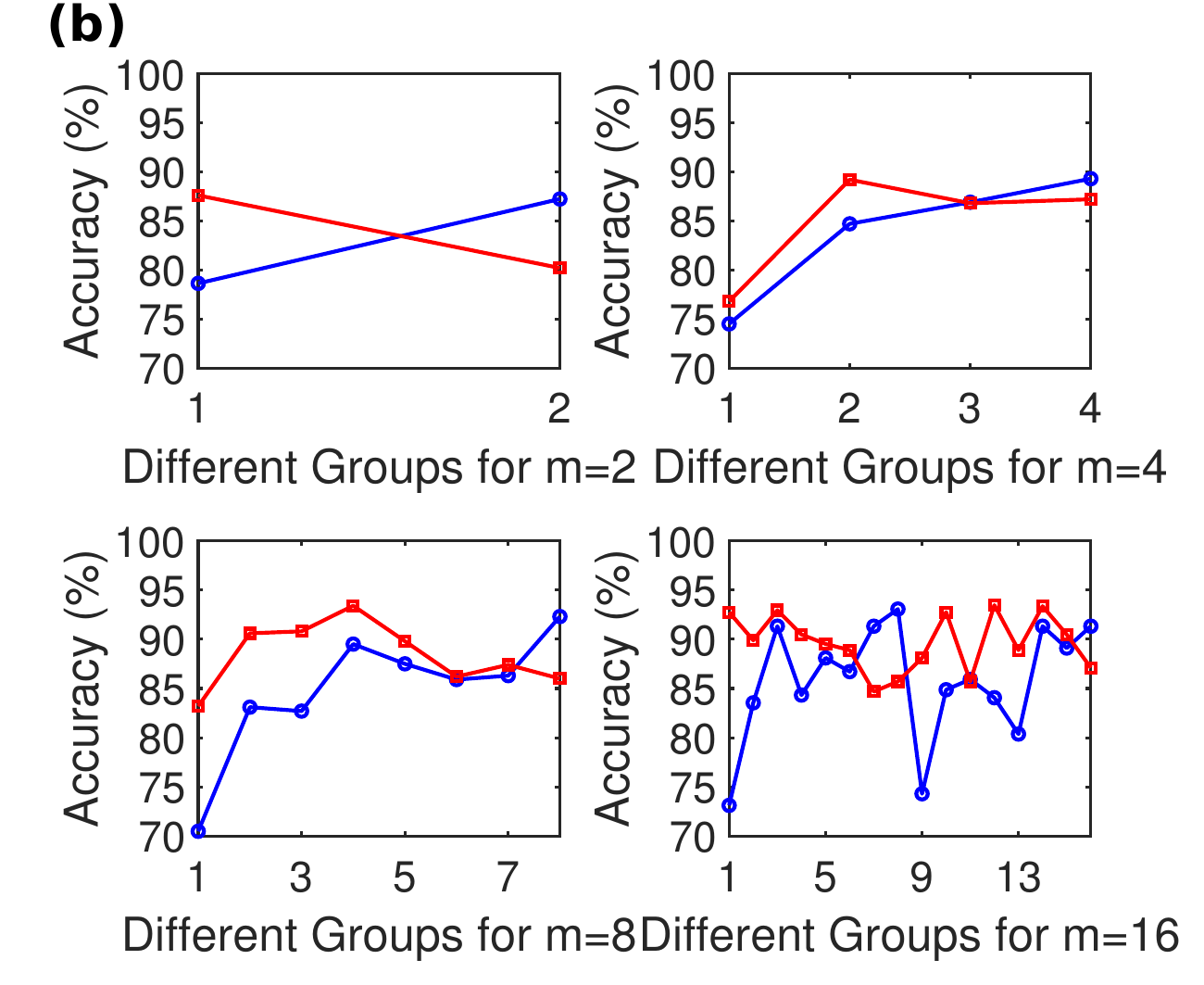}
	\end{minipage}
	\begin{minipage}[t]{0.32\textwidth}
		\centering
		\includegraphics[width=\textwidth]{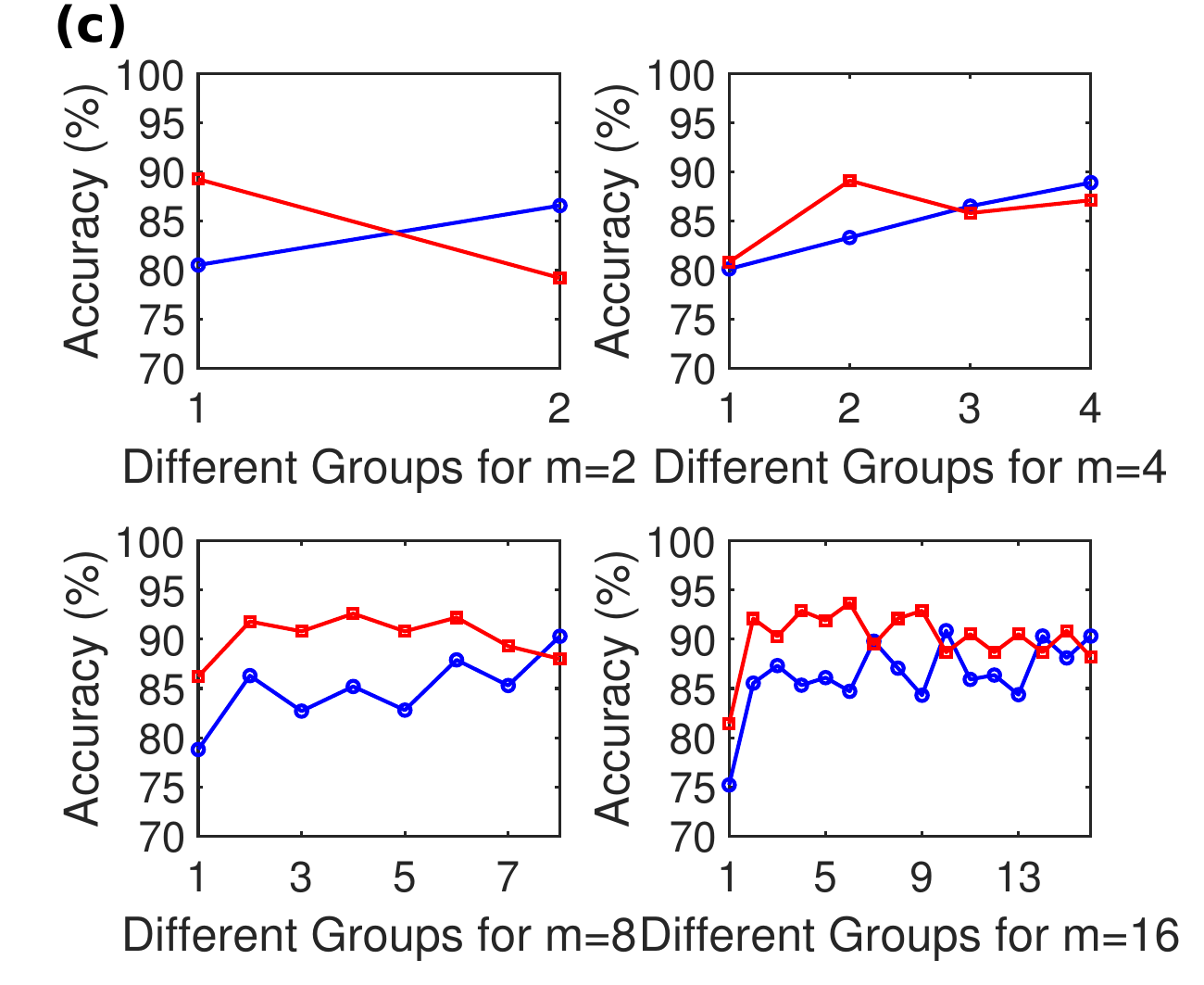}
	\end{minipage}
	\caption{When $m=2$, 4, 8, 16, average accuracy ($\%$)  of three algorithms under different $m$. The accuracies by the Renewal SVM-S4VM are represented by red lines with square, the accuracies by the SVM-S4VM are represented by blue lines with circle. (a) The prediction accuracy of each group at $l=40$. (b) The prediction accuracy of each group at $l=60$. (c) The prediction accuracy of each group at $l=80$.
	}\label{FIG.7}
\end{figure*}

In Fig.~\ref{FIG.4}, we use six different groups of 60 labeled quantum states to implement SVM-S4VM and Renewal SVM-S4VM,  obtain the classification accuracy of 2000 unlabeled quantum states in the case of $m=2$, 4, 8, 16, and compare it with the direct prediction results of S4VM without group prediction. As shown in Fig.~\ref{FIG.4}, the situation is approximately the same as that of the 40 labeled quantum states. In all four cases, SVM-S4VM and Renewal SVM-S4VM were greater than S4VM, and the prediction accuracy of Renewal SVM-S4VM was greater than that of SVM-S4VM. It is obvious that the improvement in the Renewal SVM-S4VM prediction accuracy on SVM-S4VM is more obvious when $m=8$ and 16, and the improvement is smaller at $m=2$, 4. Except for the fourth group of $m=8$, 16, SVM-S4VM hardly improves as $m$ increases, and the prediction accuracy of most SVM-S4VM and Renewal SVM-S4VM increases with the increasement of $m$.

In Fig.~\ref{FIG.5}, we use six different groups of 80 labeled quantum states, resulting in a similar result to the previous $l = 40,\ 60$, so it can be seen that the Renewal SVM-S4VM has strong universality and can effectively adapt to predict a large number of unlabeled states for a small number of labeled states.

In Fig.~\ref{FIG.6}, as the number of unlabeled quantum states increases, so does the classification accuracy. Considering that in actual use, the time required and capital cost of labeling continue to increase, but the effect improvement is small and the efficiency is not high, this paper will not continue to add more corresponding cases of labeling quantum states.

In Table \ref{tab.2}, when $l$ is 40, the average maximum prediction accuracy of the Renewal SVM-S4VM is 86.16$\%$, which is approximately 4$\%$ higher than that of the SVM-S4VM. The average maximum prediction accuracy of Renewal SVM-S4VM was 89.72$\%$ when l was 60, which was 4.25$\%$ higher for the same $l$. In summary, the accuracy of the Renewal SVM-S4VM is higher than that of the SVM-S4VM, which can effectively improve the entanglement classification accuracy in three-qubit quantum states.

\begin{table}[t]
	\renewcommand\arraystretch{1.5}
	\tabcolsep=0.205cm
	\begin{tabular}{l*{4}c}
\hline
\hline
		$l$  & S4VM & SVM-S4VM & Renewal SVM-S4VM &  \\ \hline
		40 & 78.76 & 82.17    & 86.16       &  \\
		60 & 82.30  & 85.47    & 89.72       &  \\
		80 & 82.80  & 85.91    & 90.52       &  \\ 
\hline
\hline
	\end{tabular}
	\caption{The average maximum prediction accuracy ($\%$) of the S4VM, SVM-S4VM and Renewal SVM-S4VM methods with $l=40$, $60$, and $80$.}\label{tab.2}
\end{table}

In Fig.~\ref{FIG.7}, when $m = 2$, $4$, $8$, and $16$,  the Renewal SVM-S4VM has far more data points than the SVM-S4VM, which is why the Renewal SVM-S4VM is more accurate than the SVM-S4VM. Moreover, the prediction accuracy of the first group of Renewal SVM-S4VM was higher than that of SVM-S4VM, while the prediction accuracy of the last group was lower than that of SVM-S4VM. This is because the Renewal SVM-S4VM predicts from the outside to the inside. The earliest data are distributed far from the fitting line, which is easy to classify and predict. During the iterative process, the data distribution is close to the fitting line. In contrast, it is more difficult to distinguish than the SVM-S4VM, which randomly shuffles the data group, resulting in a decrease in classification accuracy. It is also observed that during the iterative prediction process, the classification accuracy increases circuitously, which is related to the increase in the training group.

Compared with supervised learning, the accuracy of semi-supervised learning is approximately close to the accuracy of supervised learning, and even exceeds that at $n=8$, 16. This shows that Renewal  SVM-S4VM semi-supervised learning can be well applied to three-dimensional entangled state discrimination problems, which can not only reduce the training time, but also maintain good accuracy. We believe that this method can be applied for more bit entanglement discrimination of quantum states.

\section{Discussion and conclusion}\label{sec:leve4}
Semi-supervised machine learning makes good use of and extends the existing quantum state samples to guide the modeling process, and further improves the modeling performance of quantum entanglement state prediction based on supervised machine learning. In this paper, two semi-supervised learning methods, S4VM and Renewal S4VM, are used to verify the Renewal SVM-S4VM algorithm based on numerical simulation data, which highlights the superiority of the proposed method. The goal of future research is to train more complex states, combine multiple quantum state entanglement determination methods, and achieve a more efficient and stable semi-supervised method.



\begin{figure}[t]
	\centering
	\includegraphics[width=3.5in]{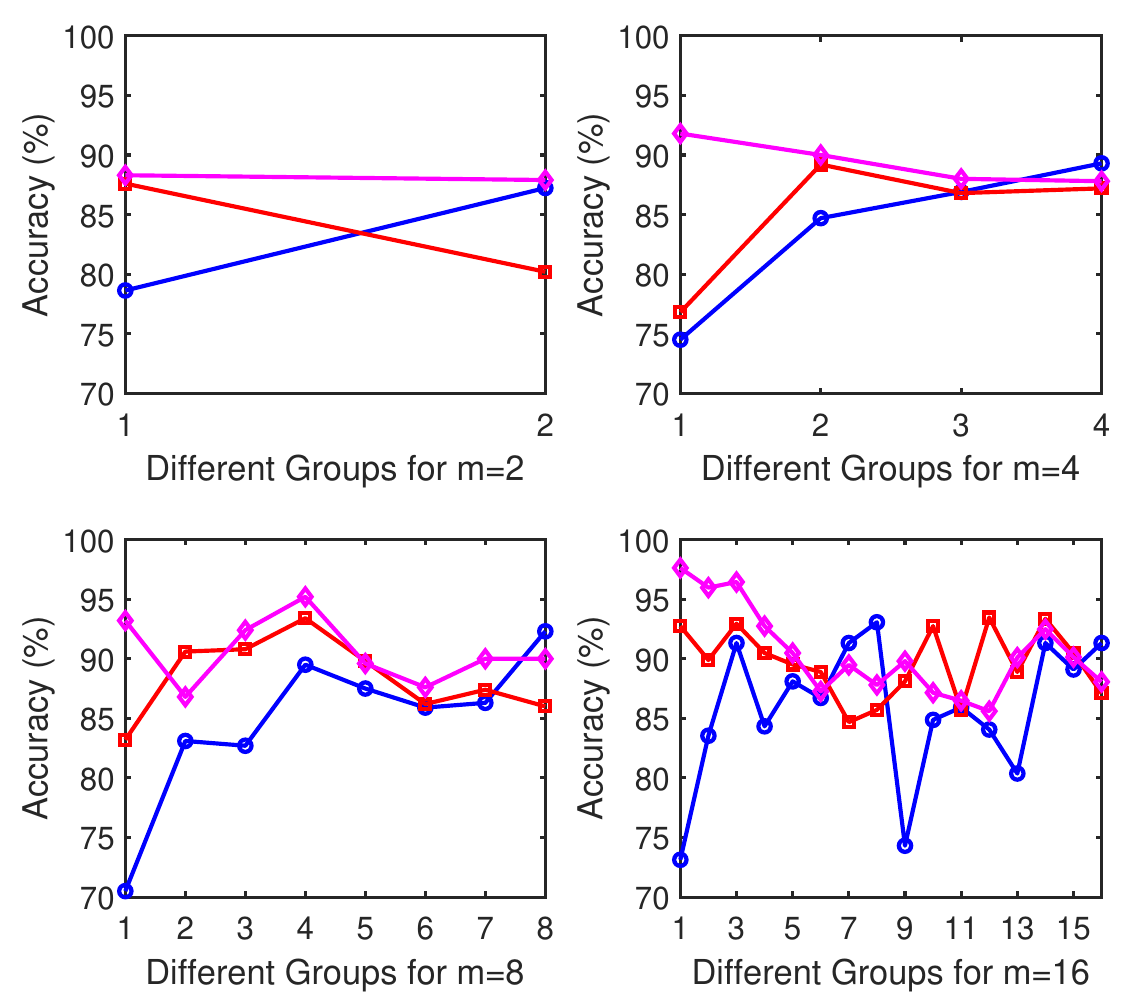}\\
	\caption{The average accuracy of the three algorithms under $l=60$. The accuracies of Renewal SVM-S4VM combined with active learning is represented by pink lines with rhombus, the accuracies by the Renewal SVM-S4VM are represented by red lines with square, the accuracies by the SVM-S4VM are represented by blue lines with circle.}\label{FIG.9}
\end{figure}


To improve prediction accuracy, we propose a method that uses active learning \cite{schohn2000less,leng2013combining,cai2015entanglement} to select the most informative 60 labeled data points from the original random set. We assume that we have 2000 labeled data points in total, and we calculate the average trace distance between each point and the rest of the points. We sort the points by their average trace distance in ascending order and pick the top 60. The idea is to choose the points that are closest to the center of the data distribution, so that the S4VM model can better distinguish the two classes of data, as shown in Fig.~\ref{FIG.9}. We find that our method can achieve a maximum accuracy of $97.62\%$. However, this also requires more labeled data. In future work, we can use generative models to augment the scarce labeled data set and meet the data demand.

Machine learning can be used to detect entangled states, but it may also misclassify non-entangled states as entangled ones, which is undesirable. Similar problems have also been reported in Refs. \cite{zhang2021einstein, roik2021accuracy,Chen2023}. To minimize this type of error,  two possible methods can be used. The first method is to select an appropriate cost function \cite{scala2022quantum} and define a hyperplane that separates some entangled states from all other non-entangled and weakly entangled states. The hyperplane can be adjusted to maximize the number of entangled states on one side. The second method is to modify the decision threshold \cite{roik2021accuracy}, which can lower the error probability. Ref. \cite{roik2021accuracy} shows that this probability can be reduced to less than $1\%$ by optimizing the threshold.

In conclusion, we explored how to use supervised and semi-supervised machine learning methods to identify genuine multipartite entanglement in three-qubit systems. We randomly generated three-qubit density matrices and trained an SVM to classify them as genuinely entangled or not. We found that using only a subset of the support vectors improved the prediction accuracy of the SVM. Furthermore, we proposed an enhanced training algorithm for S4VM, which optimizes the partitioning of the unlabeled samples and performs iterative predictions on them. Our numerical simulations confirmed that this algorithm can significantly increase the prediction accuracy of S4VM.


\section*{Acknowledgment}
This work is supported by the National Natural Science Foundation of China (Grant Nos. 11734015 and 92265209), and the Open Funding Program from the State Key Laboratory of Precision Spectroscopy (East China Normal University, Grant No.~SKLPS-KF202205), and K.C. Wong Magna Fund in Ningbo University.

\bibliographystyle{amsalpha}

\end{document}